\documentclass[prd,aps,nofootinbib]{revtex4}
\usepackage{epsfig}
\usepackage{amsfonts}
\usepackage{rotating}
\usepackage{sparticles}
\usepackage{dcolumn}
\usepackage{bm}
\usepackage{hyperref}
\hypersetup{
    bookmarks=true,         
    unicode=false,          
    pdftoolbar=true,        
    pdfmenubar=true,        
    pdffitwindow=false,     
    pdfstartview={FitH},    
    pdftitle={My title},    
    pdfauthor={Author},     
    pdfsubject={Subject},   
    pdfcreator={Creator},   
    pdfproducer={Producer}, 
    pdfkeywords={keyword1} {key2} {key3}, 
    pdfnewwindow=true,      
    colorlinks=true,       
    linkcolor=red,          
    citecolor=cyan,        
    filecolor=magenta,      
    urlcolor=green,           
    linktocpage=true
}
\usepackage{amsmath,amssymb,amsthm,graphicx,latexsym}
\usepackage{subfigure}
\usepackage{pstricks}
\usepackage{color}

\usepackage{mathrsfs}

\newcommand{\be}{\begin{equation}}
\newcommand{\ee}{\end{equation}}
\newcommand{\bea}{\begin{eqnarray}}
\newcommand{\eea}{\end{eqnarray}}
\newcommand{\bml}{\begin{subequations}}
\newcommand{\eml}{\end{subequations}}
\newcommand{\bfig}{\begin{figure}}
\newcommand{\efig}{\end{figure}}

\newcommand{\slashed}{\hspace{-1.1ex}/}

\def\bea{\begin{eqnarray}}
\def\eea{\end{eqnarray}}
\def\ba{\begin{array}}
\def\ea{\end{array}}

\def\beq{\begin{equation}}
\def\eeq{\end{equation}}

\begin{document}
$~~~~~~~~~~~~~~~~~~~~~~~~~~~~~~~~~~~~~~~~~~~~~~~~~~~~~~~~~~~~~~~~~~~~~~~~~~~~~~~~~~~~~~~~~~~~~~~~~~~~~~~~~~~~~~~~~~~~~~$\textcolor{red}{\bf TIFR/TH/16-07}
\\ \\
\title{\LARGE \textsc{\fontsize{95}{100}\selectfont \sffamily \bfseries{\blue Cosmological hysteresis in cyclic universe from membrane paradigm}}}

\author{\bf \textsc{\fontsize{10}{15}\selectfont \sffamily \bfseries{Sayantan Choudhury \footnote{\bf\textcolor{red}{\bf Electronic address: {sayantan@theory.tifr.res.in, sayanphysicsisi@gmail.com}}}}}}
\affiliation{\textsc{\fontsize{10}{15}\selectfont \sffamily \bfseries{Department of Theoretical Physics, Tata Institute of Fundamental Research, Colaba, Mumbai - 400005, India
\footnote{\textcolor{red}{\bf Presently working as a Visiting (Post-Doctoral) fellow at DTP, TIFR, Mumbai.}}}}}

\author{\bf \textsc{\fontsize{10}{15}\selectfont \sffamily \bfseries{Shreya Banerjee \footnote{\bf\textcolor{blue}{\bf Electronic address: {shreya.banerjee@tifr.res.in}}}}}}
\affiliation{\textsc{\fontsize{10}{15}\selectfont \sffamily \bfseries{Department of Astronomy and Astrophysics, Tata Institute of Fundamental Research, Colaba, Mumbai - 400005, India}}}

\begin{abstract}
{\it Cosmological hysteresis} is a purely thermodynamical phenomenon
caused by the gradient in pressure, hence the characteristic equation of state
during the expansion and contraction phases of the universe are different,
provided that the universe bounces and recollapses. During hysteresis pressure asymmetry
is created due to the presence of a single scalar field in the dynamical process. 
Also such an interesting scenario has vivid implications in cosmology when applied
to variants of modified gravity models described within the framework of membrane paradigm. Cyclic universe along
with scalar field leads to the increase in the amplitude of the cosmological scale factor
at each consecutive cycles of the universe. Detailed analysis shows that the conditions which creates
a universe with an ever increasing expansion, depend on the signature of the hysteresis loop integral $\oint pdV$ and on membrane model parameters.
\end{abstract}


\maketitle
\section{\textsc{\fontsize{10}{15}\selectfont \sffamily \bfseries{Introduction}}}
Magnetic and electric materials often show a phenomenon known as hysteresis, which corresponds to a dynamical lag between its input and output. This leads to the formation of what is commonly known as the hysteresis loop. Such a phenomenon can also be created in the cosmological world, referred to as ``{\it Cosmological Hysteresis}", in the presence of cyclic universe ( \cite{eliade,jaki,starobinsky}), which bounces and recollapses repeatedly. 

As we know, a universe with identical cycles fails to solve the main problems of Big Bang model-the requirement of very special and finely-tuned initial conditions to allow the universe to evolve to its current state. These problems are commonly known as horizon and flatness problem, where the former refers to the incapability of Big Bang model to explain the near homogeneity of CMB even  though the comoving scales entering the horizon today have been far outside the horizon at CMB decoupling, and the latter refers to the requirement of an extremely fine tuning of the curvature parameter close to zero in the early universe to explain the near-flatness observed today. However, the curvature parameter begin inversely proportional to the scale factor of the universe, a way out is to generate a cyclic universe with an increasing amplitude of the scale factor after each cycle. 

 One way to develop such a scenario is to create an inequality between the pressures at the time of expansion and contraction phases of the universe. This leads to the growth of both energy and entropy of the universe after each cycle. Tolman in his paper \cite{tolman} postulated that the presence of a viscous fluid could create such disparity between the expansion and contraction phases. This unusual approach, though resulted in an inevitable increase
in entropy, helped in solving the horizon and flatness problem due to the creation of an oscillating universe with increasing expansion maximum, hence an increasing volume $V$ of the universe, after each cycle. Thus he showed a novel way of linking thermodynamical principles to the model of cyclic universe. 

 Another novel way of creating the pressure (denoted as $p$) asymmetry is through the presence of a scalar field. Such a scenario was proposed by the authors in \cite{Kanekar:2001qd,Sahni:2012er}. Scalar field affects the dynamics in a way that maintains the symmetric nature of the equation of motion (thereby avoiding entropy production) which is an advantage over the Tolman model. Such a scenario aiding a cyclic universe leads to the production of hysteresis,
defined as $\oint pdV$ (work done on/by the scalar field during one complete cycle), during each oscillatory cycle. The loop area, hence asymmetry, is largest
in case of inflationary potentials. But the phenomenon of hysteresis is 
independent of the nature of potential. Any potential with proper minimum/minima which results in phase mixing of the scalar field during
expansion, is capable of causing the phenomenon of hysteresis. In order to avoid singularity (another serious drawback of Big Bang model),
this phenomenon is applied to models where singularity is
replaced by bounce and big crunch replaced by re-collapse or turnaround. Though there exist several other models \cite{Baumann:2014nda,Baumann:2009ds,Lyth:1998xn} which succeeded in solving the horizon and flatness problems, none of them could avoid big bang singularity.

In this paper, we have further applied the above thermodynamical phenomenon
of hysteresis to higher dimensional brane world models, Einstein Gauss-Bonnet (EHGB) brane world gravity
model and Dvali-Gabadadze-Porrati (DGP) braneworld model. Both these models are braneworld models of the universe — in which the observable universe is a four dimensional timelike hypersurface (brane) embedded in a higher dimensional (bulk) space-time, where the extra dimension can be both time-like or space-like. In both these models, one sees the effect of modifying the gravity sector in the action. These models have acquired a lot of attention at present, mainly because they arise naturally in superstring theories (which requires the existence of higher dimensions) and more so because they are falsifiable. These models have several interesting cosmological features in the early and present universe and can also explain the present cosmological scenario successfully.  Hence, in this paper, we try to test these models for cosmological hysteresis. It will be very interesting to check whether the phenomenon of hysteresis, when applied to brane world models, results in a universe with increasing expansion volume. Then such a scenario can act as an alternate solution to inflation in future, once it explains other observations like structure formations, CMB, etc.. Another advantage of studying these models in the context of hysteresis, is the presence of elements which can cause bounce and turnaround in the universe, a prime requirement in the present context. In this present work, though our analysis has been restricted to brane world models, one can always extend this scenario to other modified gravity models. For a complete analysis for other models one may refer to \cite{Choudhury:2015baa,Choudhury:2015fzb}. A notable feature
of this analysis is that an increase in expansion maximum after each cycle now
depends on the sign of $\oint pdV$ and also on the parameters of
the model. Hence applying these models to this unusual scenario (bounce+hysteresis) helps to put constraints
on the EHGB and DGP model parameters. Though the analysis that
we have performed holds good under certain physically acceptable approximations and limiting cases, we can at least show that if there are any limiting cases of the models
which can give rise to the phenomenon of cosmological hysteresis. 

Though the phenomenon of hysteresis is commonly attached to magnetic and electric systems,
there are various advantages which a cyclic universe along with the phenomenon of hysteresis within the framework of membrane paradigm, enjoys over other models:
\begin{itemize}
\item The phenomenon of hysteresis is important due to the simplicity with which it can be generated. Only a thermodynamic interplay
between the pressure and density, creating an asymmetry during expansion
and contraction phase of the universe, succeeds in causing cosmological hysteresis.
\item As the phenomena of cosmological hysteresis deals with the bouncing as well as the recollapsing phase of
the universe, one can avoid the appearance of Big Bang Singularity as well as the Big Crunch at early and late times.
\item Hysteresis can be generated by the presence of a single scalar field, which has already been
studied for a wide variety of physical situations. The most exciting issue is, it do not require any other
fields for its occurrence. Hence it is very easy to handle and its properties can be studied extensively in cosmological literature.
\item A cyclic universe having the conditions to cause hysteresis, can solve all
the Big Bang puzzles (horizon and flatness problem), hence can act as an alternative proposal to inflation, if it
explains other observations successfully.
\item In this scenario, we can always start with a closed or open universe and after allowing
the universe go through a number of cycles, we get the present observable
flat universe. In this paper, we will extensively deal with the effect of hysteresis
on such higher dimensional models that can give rise to such cyclic universe. Through analytical calculations,
we will show that irrespective of the nature of the universe, we can get a cyclic model with increasing amplitude of the scale factor.
\end{itemize}

\section{\textsc{\fontsize{10}{15}\selectfont \sffamily \bfseries{Dynamics leading to Hysteresis}}}
The scalar field equation of motion in the spatially flat FLRW cosmological background is given by:
\begin{equation} 
\ddot \phi + 3 H \dot \phi + V^{'}(\phi) = 0 \,.
\label{eq:scalar field}
\end{equation}
Here $H$ is the Hubble parameter, $\phi$ is the scalar field and $V(\phi)$ is the potential for the field $\phi$.

Let us also discuss about the origin of cosmological hysteresis scenario following:
\begin{itemize}
\item If we closely analyse Eq~(\ref{eq:scalar field}), then we find that when the universe expands i.e. $H>0$ the second term $3 H \dot \phi$ acts as friction and opposes the motion of the scalar field, thus producing a damping effect during its motion, when the universe expands ($H>0$). This lowers the kinetic energy of the scalar field compared to its potential energy, giving rise to a soft equation of state ($P=-\rho$ in case $\dot{\phi}^{2}/2<<V(\phi)$ i.e. slow roll regime).
\item By contrast, in a contracting ($H<0$) phase of the universe, when the Universe enters a contracting ($H<0$) phase, the term $3 H \dot \phi$ behaves like 
anti-friction and favours the motion of the scalar field, thereby accelerating it. This makes the kinetic energy of the scalar field much larger than the potential energy, giving rise to stiff equation of state ($P=\rho$ in case $\dot{\phi}^{2}/2>>V(\phi)$). 
\item Hence, from the second law of thermodynamics, we can convey that a net difference in the pressure (during expansion and contraction cycle) leads to a non-zero work done by/on the scalar field. In addition, if we now generate a scenario leading to the presence of bouncing and recollapsing mechanisms during contraction and expansion respectively, one can expect that a non-zero work done during a given oscillatory cycle to be converted into the energy of expansion of the universe, resulting in the growth in the maximum amplitude and hence maximum volume of the universe of each successive expansion cycle. Thus producing older and larger cycles.
\item In \cite{Sahni:2012er}, using simple thermodynamic arguments, the authors have related the change in maximum amplitude of the scale factor after successive cycles to the work done. They have shown that these equations have a universal form which is independent of the scalar field potential responsible for hysteresis. 
\item As has been discussed in \cite{Sahni:2012er}, though the process is independent of the potential, ``the presence of hysteresis is closely linked to the ability of the field $\phi$ to oscillate''. This requires the presence of potential minimum/minima in order to make the field oscillate. As a result oscillations of the scalar field during expansion randomises its phase, thereby making all values of $\dot{\phi}$ equally likely at turnaround. Thus assuring that the values of $\phi$ and $\dot{\phi}$, when the universe turns around and contracts, are nearly uncorrelated with its phase space value when the field $\phi$ began oscillating. As a result, the field almost always rolls up the potential along the different phase space trajectory compared to the one along which it had descended during expansion. 
\item This gives rise to the pressure asymmetry during expansion and contraction hence to non-zero work done, thereby leading to the  phenomenon of ``{\it cosmological hysteresis}''.
\end{itemize}
\begin{figure}[ht]
\centering
\includegraphics[width=12cm,height=8cm]{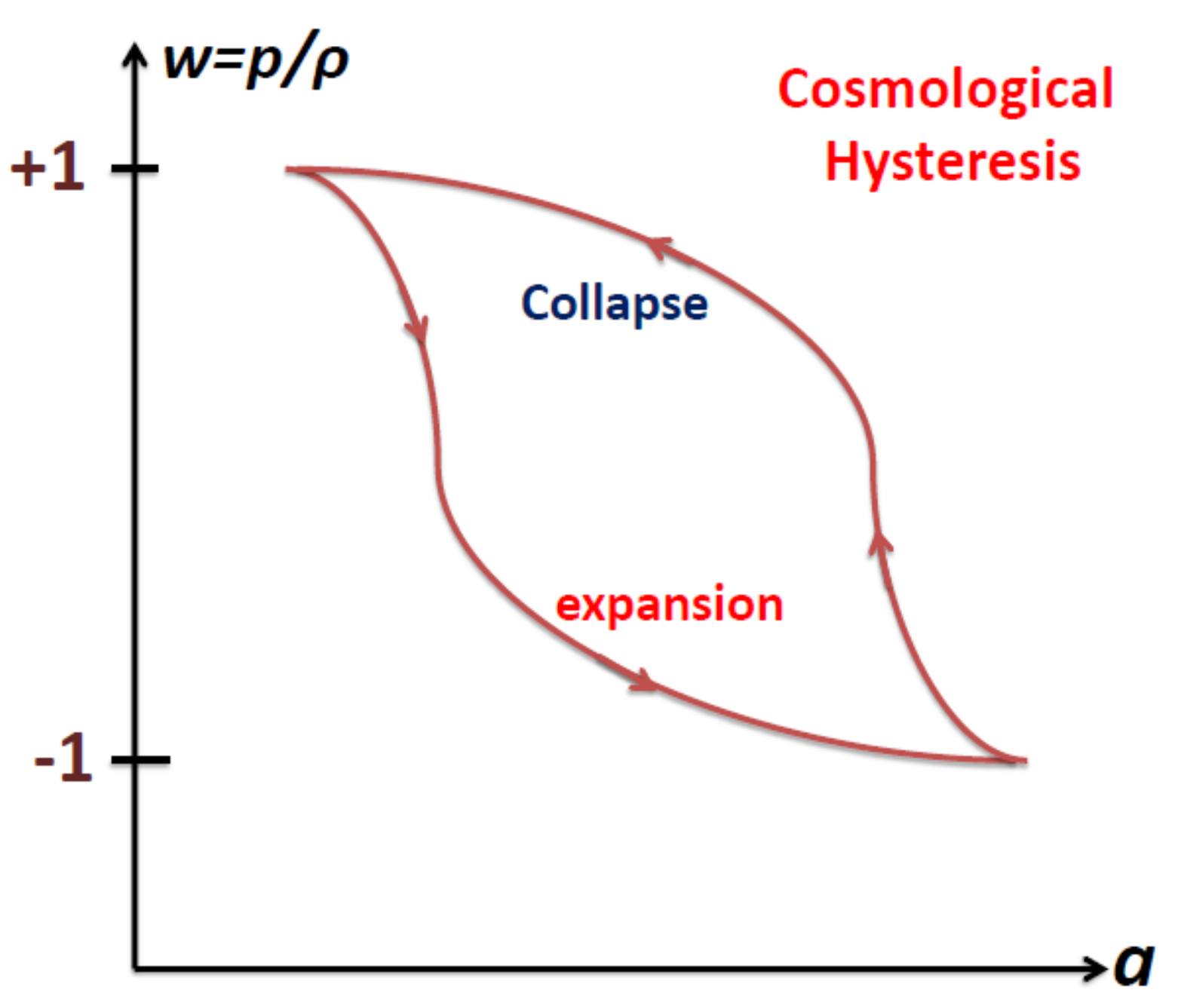}
\caption{\small An idealised illustration of cosmological hysteresis. The loop has been plotted in the $w-a$ plane.  
}
\label{fig:wa}
\end{figure}
It has been first pointed out in \cite{Kanekar:2001qd} that when we plot the equation
of state given by $w = p/\rho$ vs the scale factor $a$ from a specified cosmological model,
we get a hysteresis loop whose area contributes
to the work done by/on the scalar field during expansion and contraction of the Universe.
The general expression for the work done by/on the scalar field during one cycle is given
by \be \oint pdV = \int_{cont} pdV + \int_{exp} pdV.\ee
In the present context, the volume of the universe is given by $V=a^{3}$.
 The signature of the integral depends on the relative pressure difference  between the contraction and expansion phase i.e $ p_{cont} > p_{exp}$, then the overall signature of the $pdV$ work
 done is negative or, $\oint pdV < 0$. and vice versa. Fig.~\ref{fig:wa} shows the graphical plot between the equation of state
and scale factor of the universe. It illustrates the phenomenon of cosmological
hysteresis. For completeness also in table~\ref{fig:hysteresis}, we have mentioned the proper analogy
between cosmological 
hysteresis and magnetic hysteresis scenario.

Now following the proposal of \cite{Kanekar:2001qd,Sahni:2012er},
 we know that in order to get a cyclic universe, the condition for bounce
 and turn around are given by: \begin{enumerate}
                                \item \textbf{ Bounce}- $H=0$ and $\ddot a> 0$,
                                \item \textbf{  Turn around}- $H=0$ and $\ddot a< 0$.
                               \end{enumerate}

 \begin{figure}[ht]
\centering
\includegraphics[width=12cm,height=6cm]{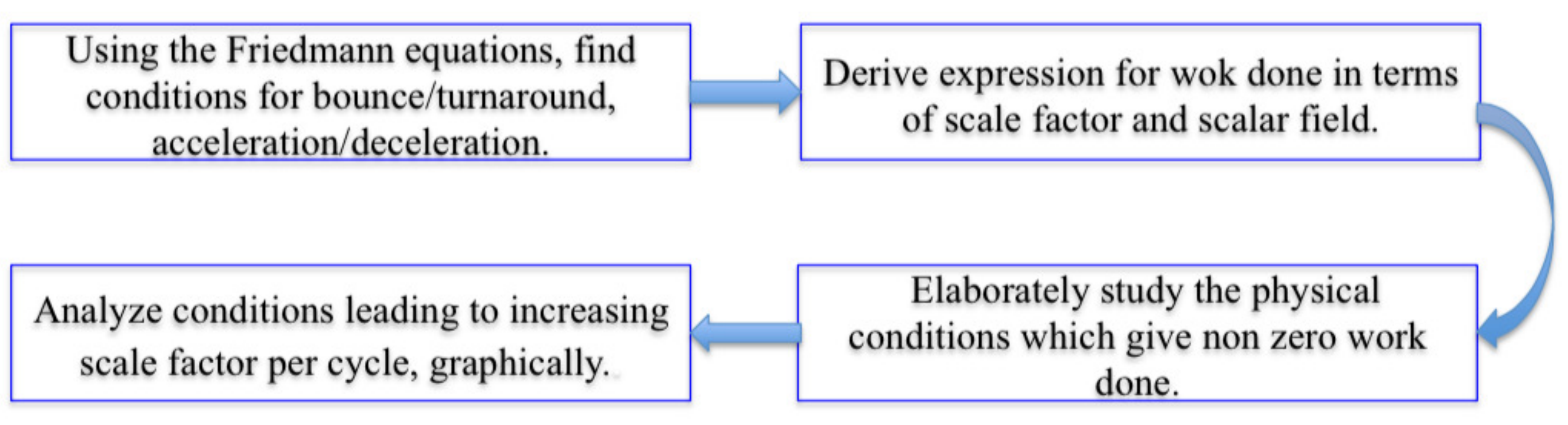}
\caption{\small A schematic representation of the basic steps in our analysis. In this paper, we have discussed the first two steps in some detail.  
}
\label{fig:algorithm}
\end{figure}

\begin{table*}
\centering
\large
\begin{tabular}{|c|c|c|
}
\hline
\hline
Characteristics & Magnetic Hysteresis & Cosmological Hysteresis
\\
\hline\hline\hline
{\bf  Loop parameters } &  $B, M$ &  $a$
\\
 &  $H$ &  $w=p/\rho$
\\
\hline
{\bf  Hysteresis loop} & $\oint H~dB$ &  $\oint w~da~(\sim \oint p~dV)$
\\
\hline
{\bf  Largest value of } & Given by hard &  Cyclic Universe\\
{\bf  hysteresis loop } &  ferromagnetic materials & with inflationary conditions
\\ \hline
{\bf  Upper and lower  } & No limit on maximum &   $w_{max}=+1$\\
{\bf  limits of the loop } &  and minimum values of $H$ & $w_{min}=-1$
\\
\hline
{\bf  Characteristic equation } & $B=\mu(H+M)$ & $w=p/\rho$ \\
{\bf  for the loop } &  & 
\\
\hline
{\bf  Nature of } & Soft ferromagnetic materials &  Universe with softer equation\\
{\bf  the loop } & have smaller loop area & of state have smaller loop area \\
\hline
\hline
\end{tabular}
\caption{ Table showing the analogy between magnetic hysteresis and cosmological hysteresis. Here $B, H, M$ represent the magnetic induction, magnetic field and the magnetisation respectively. $\rho$ corresponds to the density of the universe. Rest of the symbols used, have already been defined in the text.}\label{fig:hysteresis}
\vspace{.4cm}
\end{table*}
In this paper we will discuss the analysis leading to bounce, turnaround and hysteresis for EHGB and DGP model . Fig.~\ref{fig:algorithm} shows the basic methodology/steps for our present analysis schematically. In this paper, we have discussed the first two steps in detail for both EHGB and DGP model. The rest have been discussed in \cite{Choudhury:2015baa}.
\section{\textsc{\fontsize{10}{15}\selectfont \sffamily \bfseries{Hysteresis in EHGB braneworld}}}
The modified Friedmann equation in this model is given by (\cite{Maeda:2007cb}):
\begin{equation}
\frac{\kappa^4_5}{36}(\rho+\sigma)^2 = \left(\frac{h(a)}{a^2}+\varepsilon
H^2\right)\left[1+\frac{4\alpha}{3}
\left(\frac{3k-\varepsilon h(a)}{a^2}+2H^2\right)\right]^2\, , \label{GB-F}
\end{equation}
where $\sigma$ is the single brane tension, $\alpha$ is the Gauss-Bonnet coupling,
$\Lambda$ is the 5-D cosmological constant, $\varepsilon = +1,-1$ for space-like or time-like extra dimension respectively. Here the the function $h(a)$ is defined as \be h(a) = \varepsilon k+\frac{a^2}{4\alpha}\left(\varepsilon\mp
\sqrt{1+\frac{\alpha \mu}{a^{4}}+\frac43\alpha\Lambda}\right),\ee where $\mu$ is a constant and $k=0,\pm 1$.

Following the analysis of \cite{Maeda:2007cb}, we can rewrite the above equation in a simplified manner as:
\begin{equation} \label{plus1}
C (\rho + \sigma)^{2} = \left(A \pm H^{2}\right) \left( B + H^{2} \right)^{2} \, ,
\end{equation}
where $\pm$ corresponds to $\varepsilon = +1$ and $\varepsilon = -1$ respectively. Here $A,B$ and $C$ are defined as:
\begin{eqnarray}
A= \frac{k}{a^{2}} + \frac{h(a)}{a^{2}} \, ; B= \frac{3k}{2a^{2}} + \frac{3}{8\alpha} - \frac{\varepsilon h(a)}{2a^{2}} = \frac{3}{8\alpha} + \frac{3k}{2a^{2}}
- \frac{\varepsilon A}{2} \, ; C= \frac{\kappa_5^4}{36} \left( \frac{3}{8 \alpha} \right)^2 > 0 \,.~~~~
\label{ABC}
\end{eqnarray}

Here the necessary conditions leading to bounce and turnaround for space-like
extra-dimension have been discussed. For the analysis in the presence of
time-like extra-dimension one may refer to\cite{Choudhury:2015baa}.\\  \\
$\bullet$ \underline{\textbf{Condition for bounce:} }\\ As we know, bounce occurs when the universe
reaches its minimum radius $a_{min}$ and maximum density $\rho_{b}$. In this setup such a state of minimum radius is achieved by the universe when \be \rho_{b} = \frac{\sqrt{A}B}{\sqrt{C}} - \sigma.\ee
 The expression for change in amplitude of the scale factor at each successive cycle is given by:
\be \delta a_{min} = \frac{\oint pdV}{\left(3\sigma a_{min}^{2} - X'\right)}.\ee Here $X'$ is a new model dependent parameter which is a function of $A,B$ and $C$ (for complete expression, one may refer to \cite{Choudhury:2015baa}). Thus we see that the condition for an increase in the amplitude of the scale factor depends on $A$ through $X'$, which in turn depends on the amplitude and sign of the curvature parameter $k$ and the model parameters like $\mu,\ \alpha$ etc.
 \begin{figure}[ht]
\centering
\includegraphics[width=10cm,height=6cm]{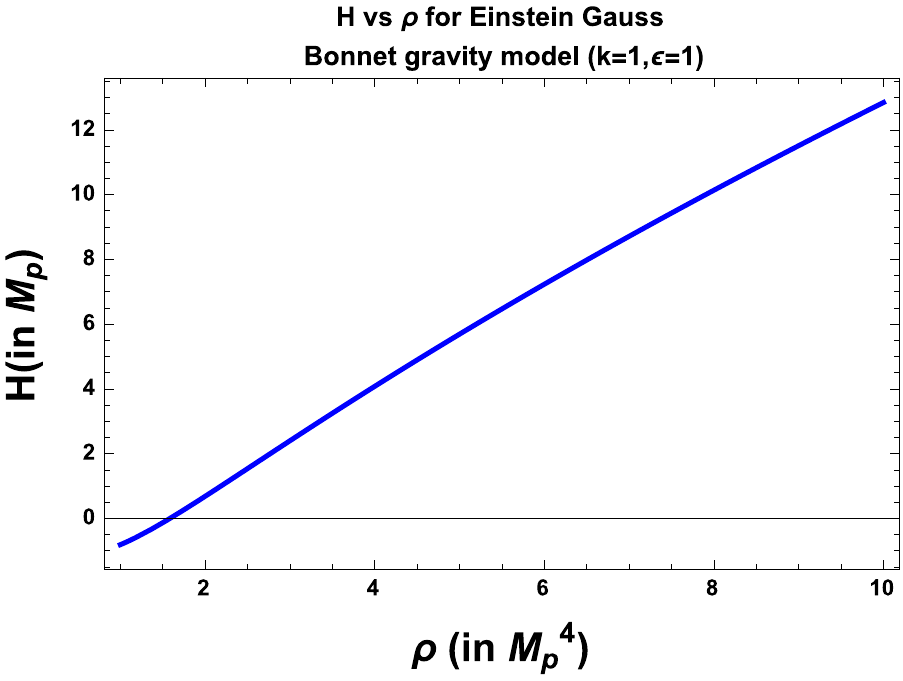}
\caption{\small An illustration of the bouncing condition for a universe with $k=1,\ \varepsilon=1,\ w=1/3,\ C=100,\ A=10M_{p}^{2},\ B=40M_{p}^{2},\ \sigma=-10^{-9}M_{4}^{4},\ C=100,\ \Lambda=6{\rm \times}10^{-3}M_{p}^{2}$.  
}
\label{fig:egb2}
\end{figure}
In Fig.~\ref{fig:egb2}, we have shown an example illustrating the phenomenon of bounce for $k= +1$ in this setup.
\\ \\
$\bullet$ \underline{\textbf{Condition for acceleration:}} \\ The second Friedmann equation in this context is given by:
 \begin{equation}
 \dot{H}+H^2=\frac{\ddot{a}}{a}  = -\frac{3\sqrt{C}^{2}(\rho + p)(\rho + \sigma)}{Y} + \left(\frac{\dot{a}}{a}\right)^{2} - \frac{Z}{Y}
 \label{gbacceleration}
 \end{equation}
 where $Z$ and $Y$ are functions of $A, B$ and their time derivatives. 
 For detailed and explicit expressions of the above quantities, one may refer to \cite{Choudhury:2015baa}. The condition for acceleration at bounce is \be p_{b} < \frac{\sqrt{A}B}{\sqrt{C}}\left(-\frac{Z}{3AB^{2}} - 1 + \frac{\sigma \sqrt{C}}{\sqrt{A}B}\right).\ee
 Thus it implies that whether the condition for acceleration violates
 the energy condition, now depends upon the values of different
 parameters of the EHGB model present in the above expression.
\\ \\ \\
$\bullet$ \underline{\textbf{Condition for turnaround:}} \\ Turnaround or re-collapse occurs when the Universe reaches its maximum radius $a_{max}$ and minimum density. Following the same line of treatment as for bounce, the condition for turnaround is \be \rho_{t} = \frac{\sqrt{A}B}{C'} - \sigma.\ee
The expression for change in amplitude of the scale factor \be \delta a_{max} = \frac{\oint pdV}{(3\sigma a_{max}^{2} - X)}.\ee Thus the conclusions for turnaround remains same as that for bounce. 
\\ \\
$\bullet$ \underline{\textbf{Condition for deceleration:}} \\ The mathematical expression reflecting the necessary  condition for deceleration is given by \be p_{t}> \frac{\sqrt{A}B}{\sqrt{C}}\left(-\frac{Z}{3AB^{2}} - 1 + \frac{\sigma \sqrt{C}}{\sqrt{A}B}\right).\ee Just like the case of acceleration within EHGB setup, the condition for deceleration
at turnaround also depends on the EHGB model parameters. 
\\ \\
$\bullet$ \underline{\textbf{Evaluation of work done in one cycle:} }

The general mathematical expression for the work done in one complete cycle (i.e contraction$\rightarrow$expansion$\rightarrow$contraction) is
\begin{equation} 
\oint pdV = \int_{a_{max}^{i-1}}^{a_{min}^i-1} 3\left(\frac{\dot{\phi}^{2}}{2} - V(\phi)\right)a^{2}\dot{a}dt + \int_{a_{min}^{i-1}}^{a_{max}^i} 3\left(\frac{\dot{\phi}^{2}}{2} - V(\phi)\right)a^{2}\dot{a}dt
\end{equation} 
where $i$ and $(i-1)$ refer to the two successive cycles i.e. $i$th and $(i-1)$th cycle of expansion and contraction phase of the Universe. Using the Friedmann equations for EHGB model, one can get the corresponding expression for work done in one cycle for this model. The complete expression has been shown in \cite{Choudhury:2015baa}, from which we can conclude that the work done now depends not only on the scale factor, but also on the different parameters of the model like the coupling constant, brane tension etc. within the present setup.

In order to get exact expressions for the evolution of the scale
factor and scalar field with time, one needs to solve the Friedmann equations under certain limiting
conditions and valid approximations. It can also be shown
that we get a non zero expression for the total work
done in one cycle under such conditions. Thus, choosing the model parameters properly, we can get an overall increase in the scale factor after one complete cycle of expansion and contraction. For more details see \cite{Choudhury:2015baa}. 

\begin{figure*}[htb]
\centering
\subfigure[ An illustration of the behaviour of the scale factor with time  during expansion phase for $\phi<<M_{p}$ with $V_{0}=10^{-8}M_{p}^{4},\ p=3$.]{
    \includegraphics[width=8.2cm,height=6cm] {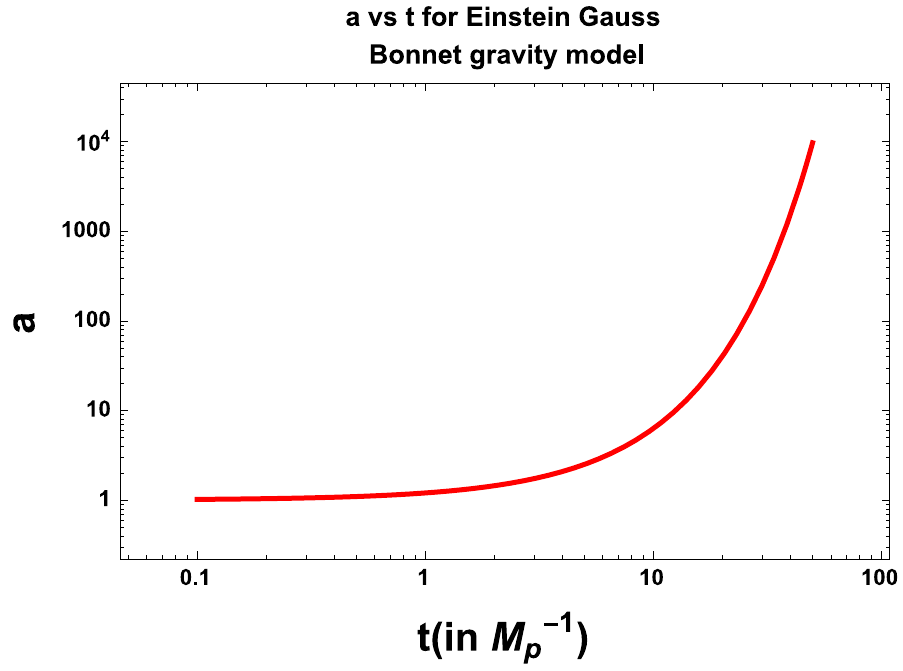}
    \label{egb7}
}
\subfigure[An illustration of the behaviour of the potential during expansion phase for $\phi<<M_{p}$ with $V_{0}=2.7{\rm x}10^{-3}M_{p}^{4},\ p=4,\ \beta=0.05$.]{
    \includegraphics[width=8.2cm,height=6cm] {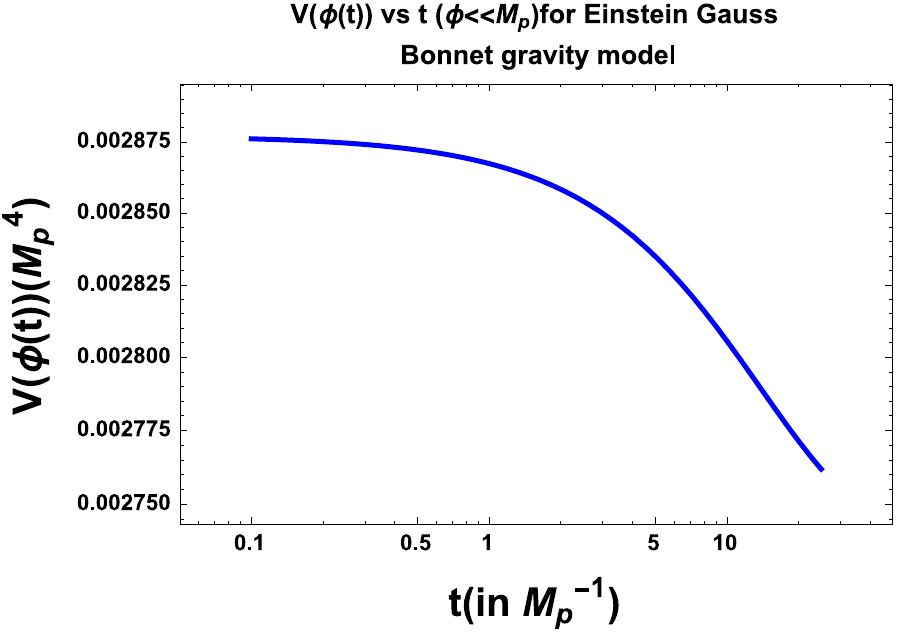}
    \label{egb8}
}
\subfigure[An illustration of the behaviour of the scale factor with time  during contraction phase.]{
    \includegraphics[width=8.2cm,height=6cm] {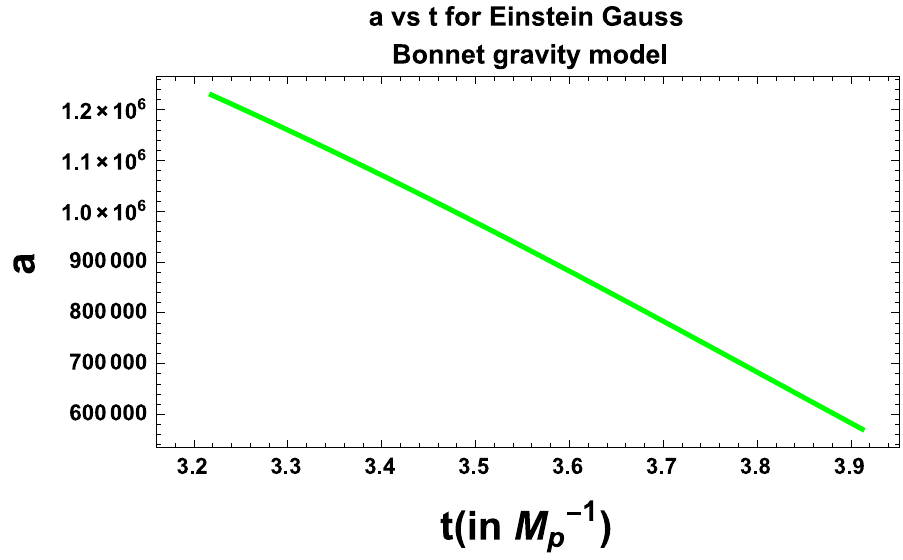}
    \label{egb10}
}
\subfigure[An illustration of the behaviour of the potential during contraction phase with $V_{0}=10^{-6}M_{p}^{4},\ p=1,\ \beta=-0.9$ .]{
    \includegraphics[width=8.2cm,height=6cm] {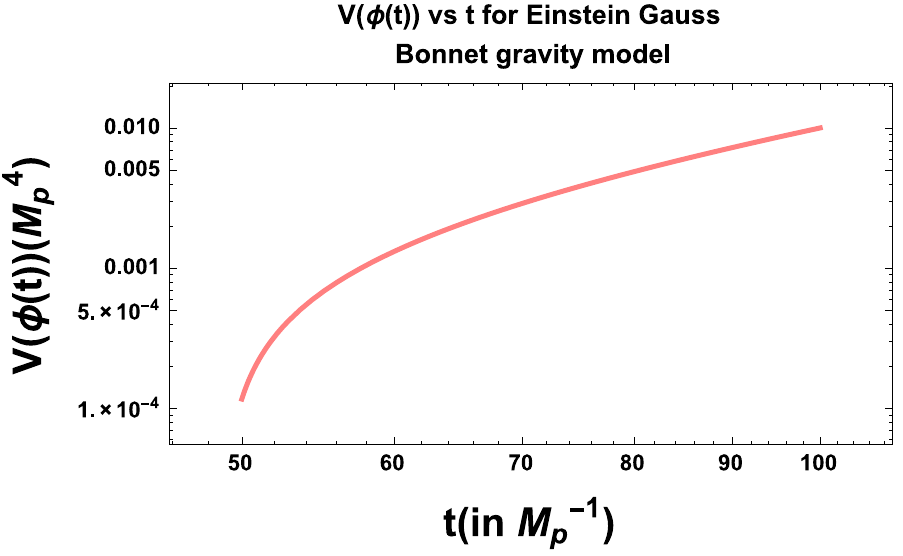}
    \label{egb11}
}    
\caption[Optional caption for list of figures]{ Graphical representation of the evolution of the scale factor and the potential during the expansion and contraction phase for Einstein Gauss Bonnet gravity model using Hilltop potential. All the graphs in this figure have been plotted in units of $M_{p}=1,\ H_{0}=1,\ c=1$, where $M_{p}$ is the Planck mass, $H_{0}$ is the present value of the Hubble parameter and $c$ is the speed of light. } 
\label{fig40}
\end{figure*}

As a verification of our claim, we have shown in Fig. \ref{fig40} the evolution of scale factor and potential during expansion and contraction phase (single cycle)
for Einstein Gauss Bonnet gravity model. Fig. \ref{fig40} has been obtained for Hilltop potential (choosing specific values for the parameters) where the form of the potential is given by \cite{Choudhury:2015jaa}:
\be\label{m64}
V(\phi)=V_{0}\left[1+\beta\left(\frac{\phi}{M_4}\right)^{p}\right]
\ee
where $V_{0}=M^4$ is the tunable energy scale and $\beta$ is the index which characterizes the feature of the potential. In principle $\beta$ can be both positive and negative.
Additionally it is important to note that, in the present context, $V_{0}$ mimics the role of vacuum energy.

Thus looking at fig. \ref{fig40}, we can clearly say that, after the universe completes one cycle, there is a net increase in the amplitude of the scale factor of the universe.

But we can always repeat the above analysis for any potential with proper minimum/minima. In \cite{Choudhury:2015baa}, we repeated the above analysis for two other forms of the potential and have found that choosing appropriate values for the model and potential parameters, we get an increase in scale factor after one cycle. 
\section{\textsc{\fontsize{10}{15}\selectfont \sffamily \bfseries{Hysteresis in DGP braneworld}}}
The modified Friedmann equation in this model is given by \cite{Copeland:2006wr}:
\bea
H^2 + \frac{k}{a^2}&=&
 \Bigg(\sqrt{ \frac{\kappa^2 \rho}{3} + \frac{1}{4r_{c}^{2}} }  + \frac{1}{2r_{c}}
 \Bigg)^2; \nonumber\\
 \dot{H} + H^2 &=&-\frac{\kappa^2}{6}(\rho + p) \left[ 1+
\left(\kappa^2 \frac{\rho}{3}   + \frac{1}{4r_{c}^2}\right)^{-1/2}
\frac{1}{2r_{c}} \right] + \left[ \sqrt{ \kappa^2
\frac{\rho}{3} + \frac{1}{4r_{c}^2}  } + \frac{1}{2r_{c}}
\right]^2.  
 \label{DGPfr1}
\eea
where \be r_{c} = \frac{M_{4}^{2}}{2M_{5}^{3}},\ee $r_{c}$ sets the scale above which the effect of extra dimension becomes important, hence is known as the crossover length scale.
\\ \\
$\bullet$ \underline{\textbf{Condition for bounce:}}\\ In the early universe, at high energy or equivalently in the high density regime of the braneworld, $\rho r_c^2/M_4^2 \gg 1$, one can expand Eq.~(\ref{DGPfr1}) as \cite{Gumjudpai:2003vv}. To a first order approximation, setting $H = 0$, we get the density at which bounce occurs as\begin{equation}
\rho_{b} = 3M_{4}^{2}\left(\frac{\sqrt{k}}{a_{b}}-\frac{1}{2r_{c}}\right)^{2}
\label{k1bounce}
\end{equation}
 The expression for change in amplitude of the scale factor at each successive cycle is now given by:
\be \delta a_{min} = -\frac{\oint pdV}{3M_{4}^{2}\left(\frac{\sqrt{k}}{a_{b}}-\frac{1}{2r_{c}}\right)a_{b}\left[3\left(\frac{\sqrt{k}}{a_{b}}-\frac{1}{2r_{c}}\right)a_{b}-2\right]}.\ee From the above expression we clearly see that, simialr to EGB model, here also the change in amplitude of the scale factor after each cycle depends on the sign of the integral, the curvature parameter and the cross over length scale for DGP braneworld. It is, though, independent of the density of the matter content of the universe.
  \begin{figure}[ht]
\centering
\includegraphics[width=12cm,height=8cm]{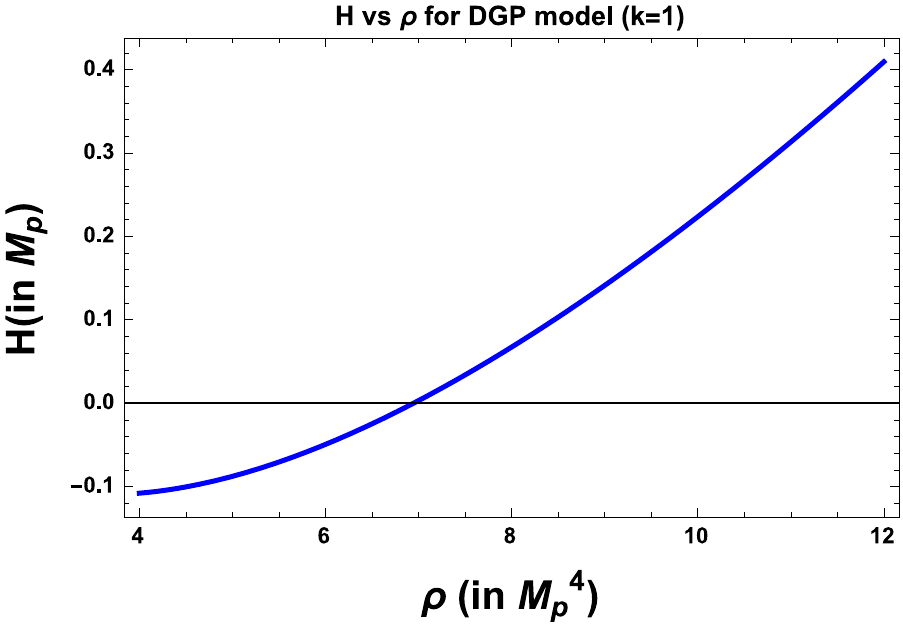}
\caption{\small An illustration of the bouncing condition for a universe with an equation of state $w=0,\ k=1,\ \mid r_{c}\mid= 1.44$.  
}
\label{fig:dgp2}
\end{figure}
In Fig.~\ref{fig:dgp2}, we have shown an example illustrating the phenomenon of bounce for $k= +1$ in this setup.
\\ \\
$\bullet$ \underline{\textbf{Condition for acceleration:}} \\ Applying once again the high energy condition to the second Friedmann equation in Eq.~ (\ref{DGPfr1}), at bounce or at high energy, we find that the contribution from $1/r_{c}$ term is small compared to the density of the scalar field and consequently one finally gets the condition
for acceleration as:
\begin{equation}
p < -M_{4}^{2}\left(\frac{k}{a_{b}}-\frac{1}{2r_{c}}\right)^{2}
\label{dgpaccel}
\end{equation}which clearly implies that the cosmological bounce can be obtained by violating the strong energy condition. Also, we see that the pressure at the time of cosmological bounce is related to the scale factor and the cross over length scale $r_c$. 
\\ \\
$\bullet$ \underline{\textbf{Condition for turnaround:}} \\ In order to get the condition for turnaround, we apply the condition, $\rho r_{c}^{2}/M_{4}^{2} << 1$, which is valid during late time universe, to the Friedmann equation. Then keeping terms upto the first order, and setting $H=0$, we get the condition for turnaround as \begin{equation}
\rho_{t} = \frac{3M_{4}^{2}}{2}\left(\frac{k}{a_{t}^{2}}-\frac{1}{r_{c}^{2}}\right)
\end{equation}
The change in the scale factor after each successive cycle is given by the following expression:
\begin{equation}
\delta a_{max} = -\frac{2}{3M_{4}^{2}\left(k - \frac{3a_{max}^{2}}{r_{c}^{2}}\right)}\oint pdV
\label{dgp6}
\end{equation}
Therefore, just like in the case of bounce, where the change in the amplitude is dependent on the parameters of the cosmological model, in such a physical prescription 
the change in the amplitude of the scale factor at turnaround also depends not only on the work done, but also on the cross over length scale $r_{c}$ of the DGP braneworld model.
\\ \\
$\bullet$ \underline{\textbf{Condition for deceleration:}} \\ Applying the late time approximation, we get the condition for deceleration as:
\begin{equation}
(\rho_{t} + 3p_{t}) > \frac{3M_{4}^{2}}{r_{c}^{2}}
\label{dgpdecel}
\end{equation}
Therefore, turnaround can be obtained without violating the energy condition.
And just like for acceleration, here we see that the condition for deceleration
depends on $r_{c}$, which was expected because in the late universe, the effect of $r_{c}$ will become more important. 
\\ \\
$\bullet$ \underline{\textbf{Evaluation of work done in one cycle:}} \\
Since the Friedmann equations in case of DGP braneworld model are highly complicated, we can use the late time and early time approximations of the Friedmann equations in order to get an physically relevant approximate analytical expression for the work done. Therefore, the work done can be decomposed into four parts as follows:
\begin{equation}
\oint pdV = \underbrace{\int_{a_{max}^{i-1}}^{a'^{i-1}} pdV + \int_{a'^{i-1}}^{a_{min}^{i-1}} pdV }_{\bf Contraction}  +  \underbrace{\int_{a_{min}^{i-1}}^{a'^{i-1}} pdV + \int_{a'^{i-1}}^{a_{max}^{i}} pdV}_{\bf Expansion}
\label{hyst2}
\end{equation}
 where the first two terms corresponds to late and early times during the period of contraction respectively and the last two
 terms corresponds to early and late times during the period of expansion respectively. Here $ a'$ corresponds to the scale factor at the time of transition $t'$ from early to late time or vice-versa,
 $a_{max}$ and  $a_{min}$ corresponds to the values of the scale factor at the time of turnaround $t_{max}$ and bounce $t_{min}$ respectively. The complete expression in terms of the model parameters has been derived in \cite{Choudhury:2015baa}, from which we can find that the evaluation of work done is independent of the parameters of this model at early times, parameter dependence enters through late time evaluation of the integral for work done.

In order to get exact expressions for the evolution of the scale
factor and scalar field with time, once again we need to solve the Friedmann equations under certain limiting
conditions and valid approximations. It can also be shown
that we once again get a non zero expression for the total work
done in one cycle under such conditions. For more details see \cite{Choudhury:2015baa}. 

\begin{figure*}[htb]
\centering
\subfigure[ An illustration of the behaviour of the scale factor with time during the early  expansion phase for $\phi<<M_{p}$ with $V_{0}=10^{-8}M_{p}^{4},\ r_{c}=0.86,\ \alpha=0.1$.]{
    \includegraphics[width=8.2cm,height=6cm] {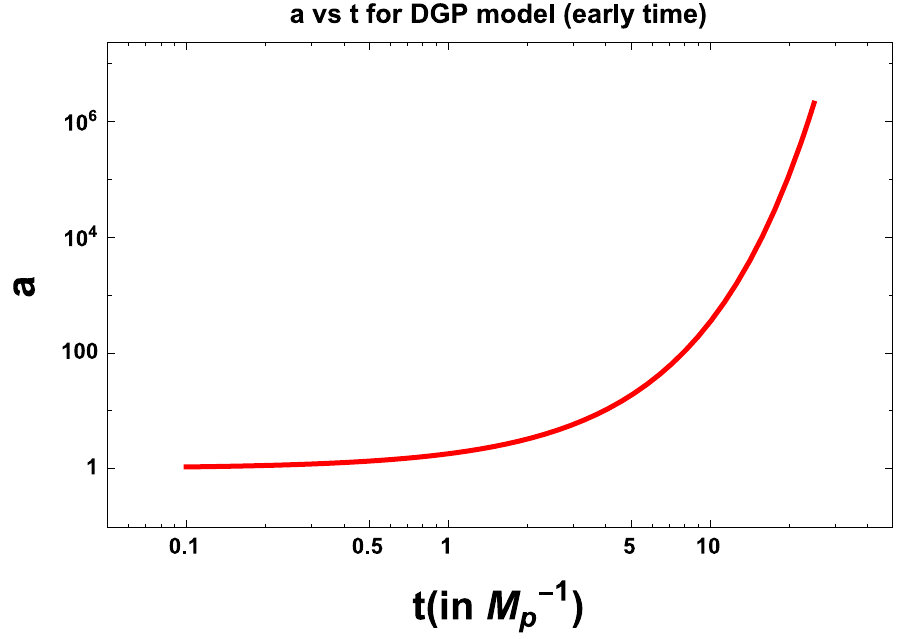}
    \label{dgp21}
}
\subfigure[An illustration of the behaviour of the scale factor with time during late time expansion phase for $\phi<<M_{p}$ with $V_{0}=6{\rm x}10^{-4}M_{p}^{4},\ r_{c}=1.66,\ \alpha=0.145,\ \beta=-1.9$.]{
    \includegraphics[width=8.2cm,height=6cm] {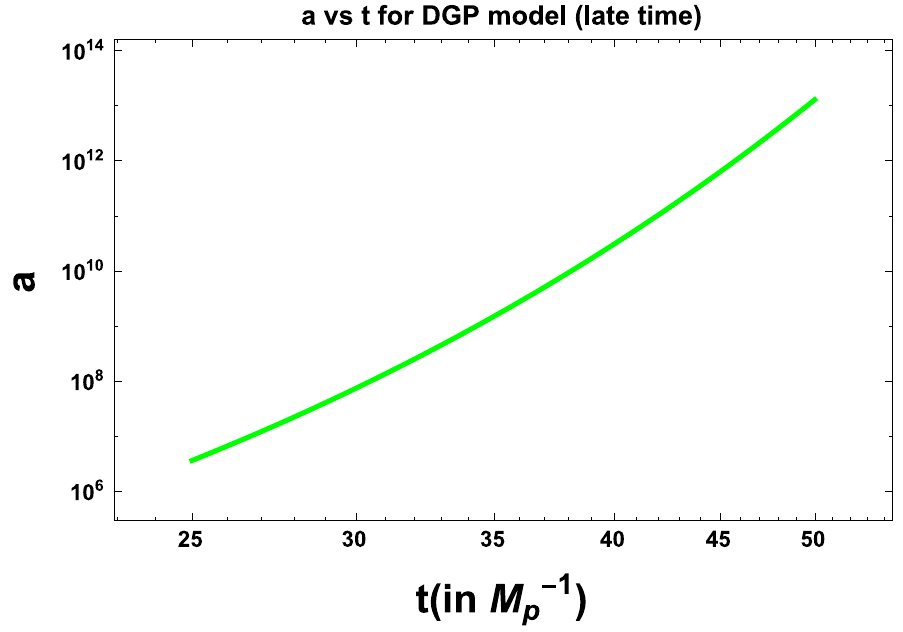}
    \label{dgp23}
}
\subfigure[ An illustration of the behaviour of the scale factor with time during the late contraction phase with $ r_{c}=6$.]{
    \includegraphics[width=8.2cm,height=6cm] {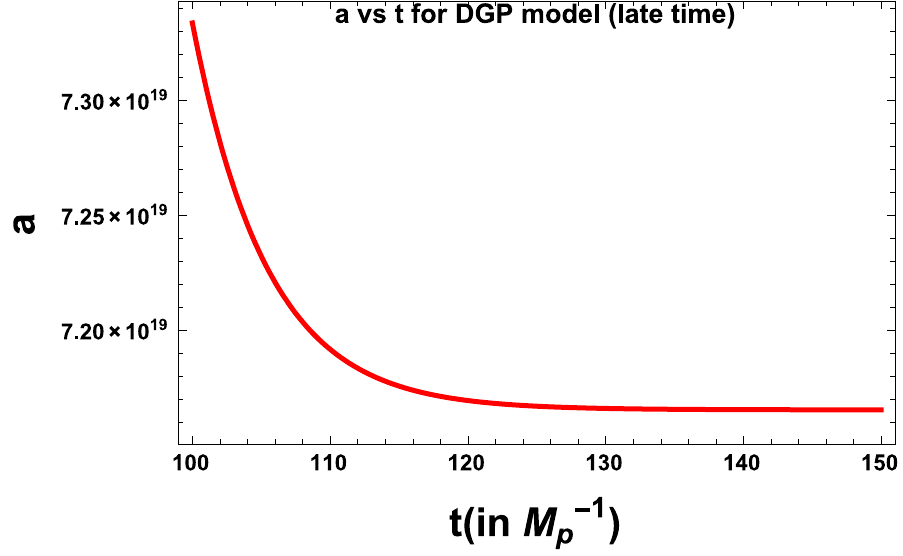}
    \label{dgp13}
}
\subfigure[An illustration of the behaviour of the scale factor with time during early time contraction phase with $\mid r_{c}\mid=2.0$.]{
    \includegraphics[width=8.2cm,height=6cm] {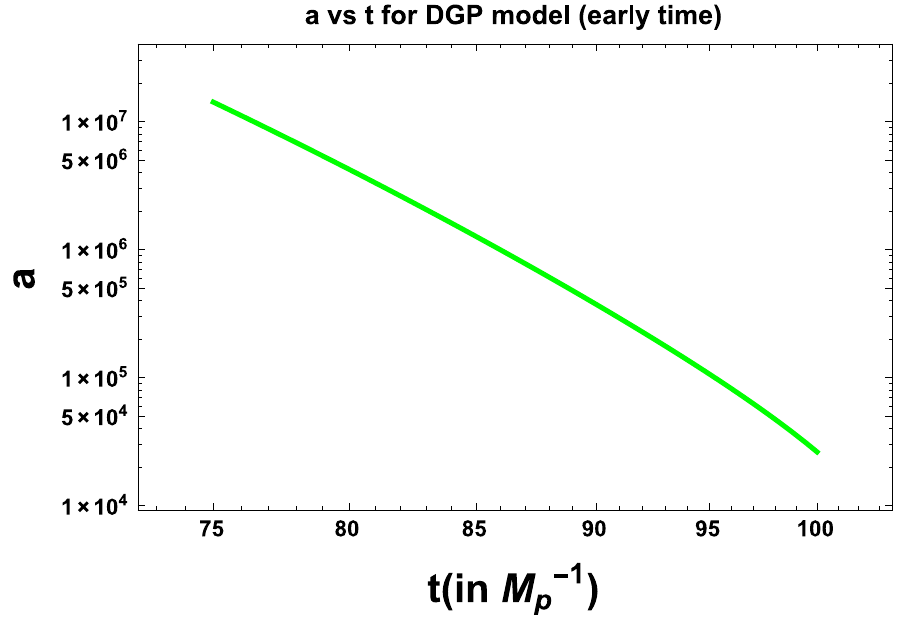}
    \label{dgp15}
}
\caption[Optional caption for list of figures]{ Graphical representation of the evolution of the scale factor during the expansion and contraction phase for the DGP model using Coleman Weinberg potential. The complete evolution of the scale factor during the expansion and contraction phases have been obtained by dividing the evolution phases into four parts (early-late expansion phase,early-late contraction phase).} 
\label{fig41}
\end{figure*}

Like for EHGB case, here also we have shown in Fig. \ref{fig41} the evolution of scale factor and potential during expansion and contraction phase (single cycle) for DGP model. Fig. \ref{fig41} has 
been obtained for supergravity motivated Coleman Weinberg potential (choosing specific values for the parameters)
where the form of the potential is given by \cite{Choudhury:2011sq,Choudhury:2011rz,Choudhury:2012ib,Choudhury:2012yh,Choudhury:2015yna}:
\be V(\phi)=V_{0}\left[1+\left\{\alpha+\beta\ln\left(\frac{\phi}{M_{4}}\right)\right\}\left(\frac{\phi}{M_{4}}\right)^{4}\right]\ee
where $V_{0}$ sets the energy scale of supergravity inflation. Additionally the model parameter 
$\alpha$ signifies the tree level effect and the parameter $\beta$ characterizes the effect of one-loop correction to the leading order result.
Here $M_{4}$ represents the background mass-scale of theory. For sake of simplicity one can consider $M_{4}$ to be the UV cut-off i.e. the Planck 
scale of the gravity theory.  

Thus just like for EHGB model, here also, looking at fig. \ref{fig41}, we can clearly say that, after the universe completes one cycle, there is a net increase in the amplitude of the scale factor of the universe.

But we can always repeat the above analysis for any potential with proper minimum/minima. Here we have shown the result for some chosen parameter values. For a more detailed description on the effect of the parameter values on these plots, one may refer to \cite{Choudhury:2015baa}.  In \cite{Choudhury:2015baa}, we repeated the above analysis for two other forms of the potential and have found that choosing appropriate values for the model and potential parameters, we get an increase in scale factor after one cycle. 


\section{\textsc{\fontsize{10}{15}\selectfont \sffamily \bfseries{Conclusion}}}
In this paper, we have explored the specific role of cosmological hysteresis scenario in the context of cyclic universe from membrane paradigm. The idea, originally proposed by the authors in refs.~\cite{Kanekar:2001qd,Sahni:2012er},
have been studied by us for higher dimensional gravity setup like DGP and EHGB model. In \cite{Choudhury:2015baa}, this study has been further extended to several other variants of modified gravity models, minimally coupled to gravity. The most interesting outcome of this analysis is the dependence
of the bouncing and turnaround
conditions on the various model parameters (specifically membrane parameters of DGP and EHGB model in this context) 
of the higher
dimensional gravity setup. Here it is important to note that, only a thermodynamic interplay between the pressure
and density, in the presence of a scalar field, succeeds in causing
cosmological hysteresis scenario.
Through this analysis we find that the phenomenon of hysteresis is very robust and is eternal for EHGB, DGP (and other) models. 
We study essentially those effective field theoretic models which can give rise to both the conditions for bounce and turnaround and at the same
time also satisfies the observed features of the present universe. Most of the results indicate that the 
value of the scale factor maximum and minimum after each cycle, depends not only on the signature of the hysteresis loop integral ($\oint pdV$) but also on the relative
amplitudes of the model parameters, or in other words, we can fine tune the model parameters and get an amplitude
increase after each cycle even if the signature of the hysteresis loop integral is positive. While doing the analysis we have
not constrained the potential by any particular form. The potential can have any general form like a power series,
oscillatory, etc, but with well defined minimum/minima which is essential for generating the required randomness or
mixing of the field in the phase space ($\dot\phi, \phi$) so that its value during contraction and expansion
are uncorrelated. This analysis also help us to constrain the parameters of the models and the structure of the
potentials, such that we get the required results compatible with observations.

The future prospects of our work are appended below pointwise:
\begin{itemize}
 \item  Through this analysis we have seen the beautiful correlation between purely
thermodynamical principle and relativistic models and how the former can be
used for extracting interesting results from the later. But the models that
we have considered are the variants of {\it minimally} coupled gravity frameworks in extra dimensional scenario. It would therefore
be interesting to investigate what new features arises once we relax this constraint.
Also we can check whether other class of modified gravity models \cite{Choudhury:2015zlc} succeeds
in generating a cyclic universe with increasing amplitude of expansion.
 
 \item One
would also like to ask what other observational signatures we can get from
such models which can be tested using CMB. In future we plan to connect these analyses with CMB observations,
by rigorous study of the cosmological perturbation theory in various orders of metric fluctuations
and computation of two point correlations to get the expressions for scalar and tensor power
spectrum in this context. 

\item We also carry forward our analysis in the development of density
inhomogeneities, which is the prime component to form large scale structures at late times. Also 
the specific role of cosmological hysteresis in the study of cosmological perturbations i.e.
for interacting/decoupled dark matter and dark energy have not been explored at all earlier. 

\item Further using the reconstruction
techniques \cite{Choudhury:2015pqa,Choudhury:2014sua,Choudhury:2014kma,Choudhury:2013iaa,Choudhury:2014wsa} one can also study the most generic features of scalar field
potentials in the framework of cosmological hysteresis.

\item Finally, it is also possible to check whether phenomenon of hysteresis can be explained through the 
effects of quantum gravity- specifically string theory \cite{Choudhury:2015hvr}, loop quantum gravity and using torsion driven frameworks \cite{Choudhury:2014hja}.
\end{itemize}

{\bf Acknowledgments:} 
 SC would like to thank Department of Theoretical Physics, Tata Institute of Fundamental
Research, Mumbai for providing me Visiting (Post-Doctoral) Research Fellowship. SC take this opportunity to thank
sincerely to Sandip P. Trivedi, Shiraz Minwalla, Soumitra SenGupta, Sudhakar Panda, Varun Sahni, Sayan Kar and Supratik Pal for their constant support
and inspiration. SC take this opportunity to thank all the active members and the regular participants of weekly
student discussion meet “COSMOMEET” from Department of Theoretical Physics and Department of Astronomy
and Astrophysics, Tata Institute of Fundamental Research for their strong support. SC also thank Sandip Trivedi and Shiraz Minwalla for giving the opportunity to be
the part of String Theory and Mathematical Physics Group. SC also thank the other
post-docs and doctoral students from String Theory and Mathematical Physics Group for
providing an excellant academic ambience during the research work. SC additionally take this
opportunity to thank the organizers of STRINGS, 2015,
International Centre for Theoretical Science, Tata Institute of Fundamental Research (ICTS,TIFR),
Indian Institute of Science (IISC) and specially Shiraz Minwalla for giving the opportunity
to participate in STRINGS, 2015 and also providing the local hospitality during the work. SC also thank 
the organizers of National String Meet 2015 and International Conference on Gravitation and Cosmology, IISER, Mohali 
and COSMOASTRO 2015, Institute of Physics for providing the local hospitality during the work. 
SC also thanks the organizers of School and Workshop on Large Scale Structure: From Galaxies to Cosmic Web, 
The Inter-University Centre for Astronomy and Astrophysics (IUCAA), Pune, India
and specially Aseem Paranjape and Varun Sahni for providing the academic visit during the work. 
Last but not the least, I would
all like to acknowledge our debt to the people of India for their generous and steady support for research in natural
sciences, especially for theoretical high energy physics, string theory and cosmology.





\end{document}